\documentclass[pra,%
 reprint,
 amsmath,assume,
 aps,
]{revtex4-2}
\usepackage{mathrsfs} 
\usepackage{graphicx}
\usepackage{dcolumn}
\usepackage{lipsum}
\usepackage{bm}
\usepackage[usenames]{color}
\usepackage{colortbl}
\usepackage{hyperref}
\usepackage{natbib}
\begin{document}

\title{
Demonstration of an optical microwave rectification by a superconducting diode with near $100\%$ efficiency}

\author{Razmik A. Hovhannisyan}
\author{Amirreza Lotfian}
\author{Taras Golod}
\author{Vladimir M. Krasnov}
\email{vladimir.krasnov@fysik.su.se}
\affiliation{ Department of Physics, Stockholm University, AlbaNova University Center, SE-10691 Stockholm, Sweden} 

\date{\today}
\begin{abstract}

Superconducting electronics offer significant advantages in speed and power efficiency for next-generation computing and communication systems. However, their practical deployment is limited by the absence of simple, efficient, and scalable superconducting counterparts to key semiconductor components. 
In this work, we investigate diodes based on planar Josephson junctions fabricated from a conventional niobium superconductor. The nonreciprocity in these diodes arises from the self-field effect induced by the geometrical asymmetry of the junction. By deliberate tuning of the junction parameters, we achieved effectively infinite nonreciprocity (within experimental resolution), characterized by a complete suppression of the superconducting critical current in one direction while maintaining a significant current in the opposite direction.
The key novelty of this work lies in the demonstration of the optical diode effect. We observed threshold-free rectification of 75 GHz microwave radiation, indicating that these diodes exhibit near-ideal optical nonreciprocity. Our results open new avenues for ultrafast superconducting electronics and lay the groundwork for wireless sub-THz signal processing.

\end{abstract}

\maketitle

\newpage


A transition to 6G wireless communication operating at upper microwave (MW) and lower terahertz (THz) frequencies presents numerous technological challenges \cite{6G_1}. Many of the requirements for 6G pose serious difficulties for traditional semiconductor electronics \cite{6G_2}, underscoring the need for innovative post-CMOS solutions. Superconducting electronics offer the potential for drastic improvements in operational speed, power efficiency, bandwidth, and noise performance \cite{Likharev_1991, Ortlepp_2014, Tolpygo_2019, Mukhanov_2019}.

The upper frequency limit of superconducting devices is determined by the 
energy gap, which is few THz for low-$T_c$ superconductors \cite{Karpov_2007, Koshelets_2025}, and extends into the full THz range for high-$T_c$ materials \cite{Kleiner_2013, Borodianskyi_2017}. Recently, THz frequency modulation using high-$T_c$ Josephson junctions (JJs) was demonstrated \cite{Kakeya_2024}, highlighting the potential of superconducting electronics for THz signal processing.
However, the practical realization of this concept requires a complete suite of THz superconducting components—including digital devices, local oscillators, mixers, demodulators, and more. One of the main obstacles in this direction is the lack of technologically simple, operationally efficient, and submicron-scalable analogs for key semiconductor components.

The diode is one of the primary electronic building blocks. Diodes find diverse applications in signal processing (rectification, demodulation, etc.) as well as in digital components (logic gates, multiplexers, etc.). In recent years, the development of superconducting diodes has emerged as an active and rapidly evolving area of research \cite{Krasnov_1997,Ando_2020,Wu_2022,Lin_2022,Baumgartner_2022,Golod_2022,Kim_2023,Franke_2023,Nadeem_2023,Gupta_2023,Finkelstein_2023,Robinson_2023,Sundaresh_2023,Nagaosa_2024,Guarcello_2024,Deskhmukh_2024,Banerjee_2024,Giazotto_2025,Goldobin_2024,Qi_2025,Ono_2022,Giazotto_2022,NbSe2_antena,Moodera_2024,Berggren_2024,Gulian_2023}.  
The superconducting diode is characterized by the nonreciprocity of critical currents, $A=\mid I_c^+/I_c^-\mid$, and the diode efficiency, $\eta=(I_c^+ - \mid I_c^-\mid )/(I_c^+ + \mid I_c^-\mid )$.

The emergence of nonreciprocity requires the simultaneous breaking of both spatial and time-reversal symmetries. Spatial symmetry can be broken by utilizing noncentrosymmetric superconductors~\cite{Nagaosa_2024, Ando_2020, Wu_2022, Lin_2022, Nadeem_2023, Banerjee_2024} and heterostructures~\cite{Ono_2022, Baumgartner_2022, Sundaresh_2023}, or by introducing geometrical asymmetry in vortex ratchets~\cite{Vilegas_2003, Moschalkov_2006, Lustikova_2018, Koelle_2000, Robinson_2023} and JJs~\cite{Krasnov_1997, Golod_2022, Beck_2005, Guarcello_2024, Goldobin_2024}.
The geometrical approach allows the use of conventional superconductors, making it suitable for practical applications. Niobium-based Josephson diodes with $A \simeq 4$ at zero field and $A \simeq 10$ at finite fields have been demonstrated~\cite{Golod_2022}. Nonreciprocity in such diodes arises from the self-field effect—the back-action of the current-induced magnetic field on the highly field-sensitive $I_c$ of the JJ~\cite{Krasnov_1997, Golod_2022}.

The nonreciprocity of $I_c$ enables rectification of ac transport current. 
However, it does not guarantee wireless (optical) rectification. 
Generally, the self-field effect is caused by asymmetric current flow near or within the junction \cite{Krasnov_1997}, see the Supplementary. In the transport case, it can be directly introduced using an asymmetric bias configuration \cite{Krasnov_1997,Golod_2022,Guarcello_2024}. In the optical case, controlling the THz current distribution is less straightforward. Due to the large wavelength, $\lambda_0 \sim$ mm, the electromagnetic field is effectively uniform on the $\mu$m-scale of the JJ. Furthermore, the size mismatch inhibits effective absorption of radiation by the JJ \cite{Krasnov_2023}. Absorption is instead enabled by mm-scale electrodes acting as THz antennas. The asymmetry of the optically induced current 
is governed by the asymmetry in THz impedance, which is determined by the mm-scale antenna geometry {\em far outside} the JJ. Therefore, transport and optical diode effects are not equivalent. While rectification of transport current is inherent to all superconducting diodes 
\cite{Beck_2005,Lustikova_2018,Moschalkov_2006,Vilegas_2003,Goldobin_2024,Qi_2025,Giazotto_2025,Deskhmukh_2024,Finkelstein_2023,Gupta_2023,Kim_2023,Baumgartner_2022,Wu_2022,Krasnov_1997,Golod_2022,Guarcello_2024,Ono_2022,Giazotto_2022,NbSe2_antena,Krasnov_1997,Golod_2022,Gulian_2023,Berggren_2024,Moodera_2024}, to our knowledge, there have been no reports of high-frequency optical rectification.

\begin{figure*}[t]
    \begin{center}
    \includegraphics[width = 0.95\textwidth]{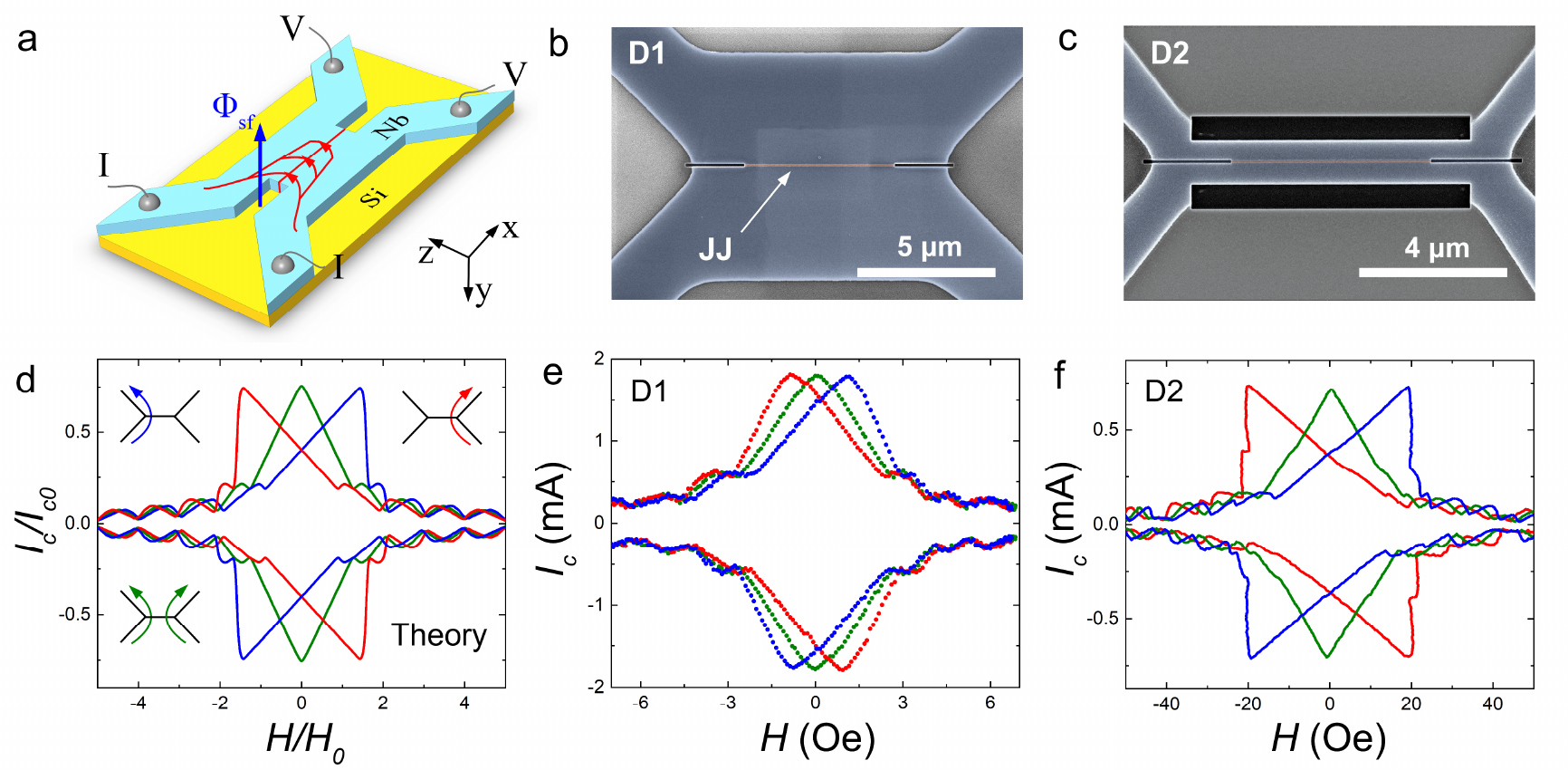}
    \caption{\textbf{Effect of bias, geometry and junction parameters on nonreciprocity.}
\textbf{a} A sketch of four-terminal Josephson diode. Asymmetric application of a bias current at one edge of the junction induces a self-field flux $\Phi_{sf}$. 
\textbf{b} and \textbf{c} SEM images of D1 and D2 devices with wide and narrow electrodes. Planar junctions are marked by orange lines. 
\textbf{d} Calculated $I_{c}(H)$ for a symmetric (olive) and asymmetric bias from left (blue) or right (red) edges. The self-field effect tilts the $I_c(H)$ patterns and induces nonreciprocity at finite field. 
\textbf{e} and \textbf{f} Measured $I_c(H)$ of D1 and D2 at $T=5$ K for three bias configurations as in \textbf{d}. Note that D2 exhibits much larger self-field effect than D1, despite a smaller $I_c$.
}
    \label{fig:1}
    \end{center}
\end{figure*}

In this work, we study diodes based on multi-terminal planar Nb junctions with self-field-induced nonreciprocity~\cite{Krasnov_1997,Golod_2022}. Our aim is twofold: First, through careful device optimization, we investigate the limit of achievable transport nonreciprocity. Second, we study optical rectification at $f=75$ GHz in the upper V/W MW bands.

Figures~\ref{fig:1}(a–c) show a sketch and scanning electron microscope (SEM) images (false color) of two (D1 and D2) of more than ten studied devices. The diodes were fabricated from a single Nb film and contain a planar JJ with two pairs of electrodes separated by narrow notches. 
The junction itself is symmetric; however, the multi-terminal geometry enables asymmetric biasing, which leads to the appearance of self-field, as sketched in Fig.~\ref{fig:1}(a).

Fig.~\ref{fig:1}(d) shows calculated $I_c(H)$ patterns for the three bias configurations sketched in the insets. The field is normalized by the flux quantization field, $H_0$. It can be seen that symmetric biasing (olive) leads to reciprocal $I_c(H)$. Asymmetric biasing from one edge leads to the appearance of a self-field, which tilts the $I_c(H)$ patterns and causes nonreciprocity at finite fields~\cite{Krasnov_1997,Golod_2022,Goldobin_2024}. The sign of self-field depends on the bias configuration. 

Figs.~\ref{fig:1}(e) and (f) show the measured $I_c(H)$ patterns for D1 and D2 at $T = 5$~K in the three bias configurations. As seen from Figs.~\ref{fig:1}(b) and (c), D1 has wide electrodes ($W_z = 3.86~\mu$m) and a junction with length $L_x = 4~\mu$m and a large linear current density $J_c$. D2, in contrast, has much narrower electrodes ($W_z = 0.48~\mu$m) and a longer junction ($L_x = 5~\mu$m) with a lower $J_c$.  
Apparently, the self-field effect in D2 is much more pronounced than in D1, indicating that nonreciprocity depends on electrode geometry and junction characteristics.

The largest nonreciprocity is achieved when the maximum of $I_c(H)$ on one side meets the minimum on the opposite side. The primary maximum in the central lobe, $I_{c0}$, corresponds to zero net flux in the JJ. For asymmetric bias, this maximum occurs at a finite external field, which is needed to compensate for the self-field flux, $\Phi_{sf}$. The first minimum occurs at one flux quantum, $\Phi = \Phi_0$. Since $I_c(H) \simeq 0$ at this point, the self-field is negligible and the flux is created solely by the external field.  
Thus, the condition for achieving maximum nonreciprocity is
\begin{equation}
    \Phi_{sf} = L_{sf} I_{c0} = \Phi_0,
\end{equation}
where $L_{sf}$ is the self-field inductance~\cite{Golod_2022}.

\begin{figure*}[!ht]
    \begin{center}
    \includegraphics[width = 0.95\textwidth]{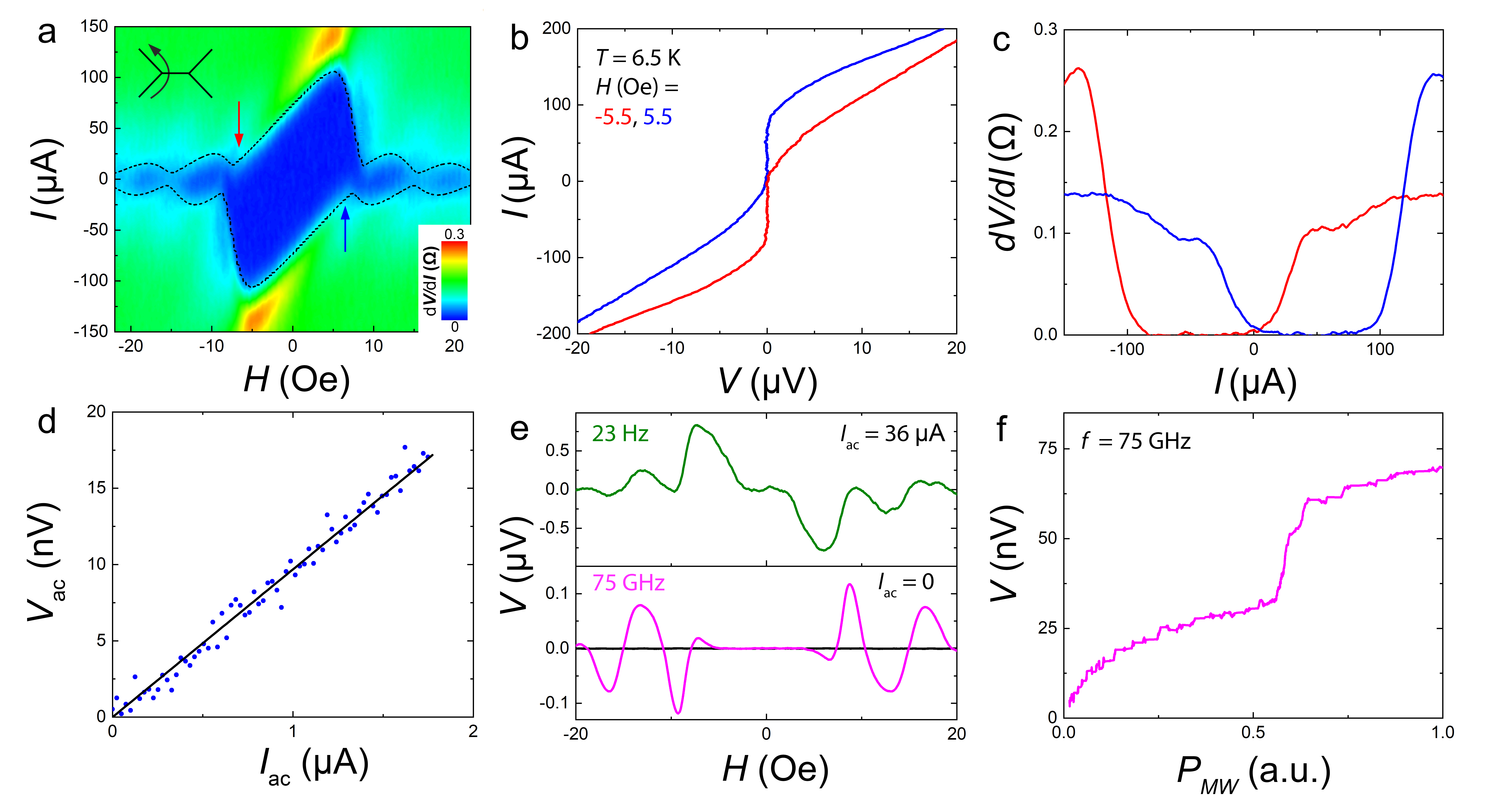}
    \caption{\textbf{Optimization of the D2 diode and demonstration of transport and optical rectification in the left-edge bias configuration at $T=6.5$ K.}
\textbf{a} The measured $I_c(H)$ modulation (differential resistance). The black dotted line represents a numerical fit for $L_x=3\lambda_J$. \textbf{b} The $I$-$V$s at $H = \pm 5.5$~Oe, indicated by arrows in (a). 
\textbf{c} Differential resistances for the $I$-$V$s from (b). Note that the critical current is significant, $\sim 100~\mu$A, in one direction, but vanishing in the opposite direction. 
\textbf{d} High-resolution lock-in measurements at $H=-5.5$ Oe. The Ohmic behavior 
confirms the absence of critical current. \textbf{e} Magnetic field dependence of the rectified dc voltage. Top panel, rectification of the low-frequency transport current. 
Bottom panel, rectification of the optical microwave radiation at $f= 75$ GHz. 
\textbf{f} The rectified dc voltage versus optical MW power.   
} 
    \label{fig:2}
    \end{center}
\end{figure*}

Both parameters in Eq.~(1) are tunable.  
$L_{sf}$ depends on the electrode geometry near or within the JJ (see Supplementary). The geometric part is determined by the notch, and the kinetic part by the $x$-component of the current density at the junction interface, which scales as  
$L_{sf} \propto L_x / W_z$.  
For D1, $L_x / W_z \simeq 1$, while for D2 it is $\simeq 10$, explaining the significant difference in the observed self-field effect.  
The critical current is given by $I_{c0} = J_c L_x$, where $J_c$ depends on temperature. According to Eq.~(1), the self-flux increases with increasing $I_{c0}$, but there is a constraint. A perfect diode ($A = \infty$) requires vanishing $I_c(\Phi_0) = 0$, which is achievable only in the short junction limit~\cite{Owen},  
\begin{equation}
L_x < 4\lambda_J,
\end{equation}  
where $\lambda_J(J_c)$ is the Josephson penetration depth. This imposes upper limits on both $L_x$ and $J_c$.

The D2 device in Fig.~\ref{fig:1}(f) is characterized by a large self-field, with the central lobe displacement by more than $3H_0$. However, the nonreciprocity is modest, $A \simeq 7$, for two reasons: the maxima are not aligned with the opposing minima, and the minima do not vanish because the JJ is in the long limit, as indicated by the broad triangular central lobe of $I_c(H)$~\cite{Owen}.  
Optimization of the diode can be done both by proper geometric design ($L_{sf}$, $L_x$) and tuning of the junction parameters ($I_{c0}$) to satisfy Eqs.~(1) and (2).

Fine-tuning of D2 was performed by raising the temperature. Figure~\ref{fig:2}(a) shows the $I_c(H)$ pattern at $T = 6.5$~K. The $I_{c0}$ was reduced, bringing the junction into the short limit, $L_x = 3\lambda_J$, as follows from the numerical fit (black line). Importantly, the central maxima are now aligned with the opposing first minima. 
Figs.~\ref{fig:2}(b) and (c) show the current-voltage ($I$–$V$) and differential resistance $dV/dI(I)$ curves at corresponding fields, $H = \pm 5.5$~Oe. It is seen that $I_c$ is significant ($100~\mu$A) in one direction but undetectable in the opposite direction. To evaluate the accuracy of the $I_c$ cancellation, we performed accurate lock-in measurements, presented in Fig.~\ref{fig:2}(d).  
The offset-free Ohmic behavior indicates that there is no critical current within our experimental resolution ($\sim 0.1~\mu$A).  
Thus, conscious optimization led to the realization of a perfect (within resolution) diode with $A > 1000$ ($100~\mu\text{A}/0.1~\mu\text{A}$), and $\eta > 99.8\%$.

Signal processing is one of the primary applications of diodes.  
The top panel in Fig.~\ref{fig:2}(e) demonstrates rectification of a low-frequency transport current with $f = 23$~Hz and amplitude $I_{ac} = 36~\mu$A. Upon variation of the magnetic field, the rectified voltage changes sign in direct correlation with the nonreciprocity of the $I_c(H)$ pattern in Fig.~\ref{fig:2}(a)~\cite{Krasnov_1997}. The magnitude of the rectified dc voltage can be directly obtained from the shape of the transport $I$–$V$ curves~\cite{Golod_2022} (see the Supplementary for details). 
However, as emphasized in the introduction, transport and optical rectification are fundamentally different: transport nonreciprocity is determined by bias asymmetry {\em close to or within} the JJ, while optical nonreciprocity arises from impedance asymmetry governed by mm-scale geometry {\em far away} from the JJ. 

To verify high-frequency optical rectification, we performed MW experiments at $f = 75$~GHz (the boarder of V/W bands). The bottom panel in Fig.~\ref{fig:2}(e) shows the magnetic field dependence of the rectified optical signal at constant MW power, $P_{MW}$. It can be seen that the magnetic field dependencies of transport (top) and optical (bottom) rectification are qualitatively similar but not identical.  
Fig.~\ref{fig:2}(f) shows the rectified dc voltage as a function of MW power at $H = -5.5$~Oe. 
It is seen that rectification starts without a threshold, indicating that not only transport but also optical nonreciprocity is close to perfect.  
The two measurements in Fig.~\ref{fig:2}(e) were performed under identical conditions; the current source was connected to the left-edge electrodes, but no transport current was applied in the optical case. Flipping the current connections from the left to the right edge of the JJ reverses the polarity of both transport and optical rectification (see the Supplementary).
 
The seeming agreement between the transport and optical diode effects is caused by the dual role of the current source. In the transport case, it directly sets the current flow asymmetry in the JJ, as sketched in Fig.~\ref{fig:1}(a). In the optical case, the low input resistance of the current source shunts one side of the MW antenna, thereby introducing the required impedance and MW current asymmetry. As a result, the effects appear similar, even though the underlying mechanisms are different.

How significant is the race for high nonreciprocity—does the diode need to be perfect? The answer depends on the specific application. For logic gates and related digital components, a modest $A \sim 2$–$4$ may be sufficient. However, large nonreciprocity is crucial for high-frequency signal processing because an imperfect diode would exhibit a threshold power for signal detection.
Moreover, a perfect diode could address one of the main challenges in superconducting circuits: the lack of effective switches for signal routing and isolation. In semiconducting electronics, a single FET transistor can open or close a line for signal transfer. In superconducting electronics, no such switch exists, leading to massive current leakage across all interconnections.
A perfect superconducting diode could enable signal and current routing, which would be indispensable for building a superconducting computer.

To conclude, the formulated conditions for achieving optimal diode performance, Eqs.~(1) and (2), combined with proper geometrical design have led to the realization of a perfect (within resolution) diode. 
We have, for the first time, demonstrated optical rectification of upper microwave (V/W-band) signal.
The demonstrated small, robust, technologically simple, and highly efficient diodes, based on conventional Nb superconductors, are suitable for practical applications. Furthermore, their polarity is switchable either by changing the bias configuration (Figs.~1d–f) or the magnetic field (Figs.~2b,e). Moreover, as shown in Ref.~\cite{Golod_2022}, polarity switching in such diodes can be achieved through controllable manipulation of a single Abrikosov vortex, which also enables operation at zero external field.
We anticipate that such diodes could find applications in future wireless communication systems and as a new class of passive signal routers in complex superconducting circuits.

\section*{Methods}
{ \bf Device fabrication}. The diodes were fabricated from a single thin (70 nm) niobium film, deposited by DC magnetron sputtering. 
The electrodes were patterned by photolithography and reactive ion etching. The variable thickness bridge-type planar JJs were made by Ga$^+$ focused ion beam (FIB) etching (a single line cut). 
The junction linear current density depends on cut depth.  
For D1 and D2 devices nominal depths were 100 and 120 nm, respectively, 
leading to significantly smaller $J_c$ for D2. 

At each side of the JJ, narrow notches separating JJ electrodes were made. The self-field is generated in the notches and the narrow width $\sim 100~$nm of the notches increases the geometric self-field inductance. The quantitative estimation of geometric and kinetic contributions to $L_{sf}$ is provided in the Supplementary.  

To determine the optimization strategy, several batches were made, each containing several diodes with different junction parameters and electrode geometries. In total more than ten diodes were analyzed in this study (see the Supplementary). Some diodes were post-processed by FIB to modify electrode geometry. Such modification is seen as two dark rectangles (FIB-etched areas) in the SEM image of D2, Fig. 1(c).      

{\bf Experimental methods}. 
The measurements were performed in a closed-cycle optical cryostat. The magnetic field perpendicular to the film was supplied by a superconducting solenoid. 
All $I$-$V$s presented in this manuscript are non-hysteretic.
High-precision lock-in measurements, Fig. \ref{fig:2} (d), were performed 
using a 30 s averaging time.

The microwave signal at $f=75$ GHz was generated by a linearly polarized frequency multiplier. The MW power was attenuated using a grid polarizer and monitored by a Golay cell detector. The MW beam was guided quasi-optically to the diode via optical windows with low-temperature Zitex filters using high-density polyethylene lenses. The rectified dc voltage was obtained by lock-in measurement at a chopper frequency. Additional clarifications can be found in the Supplementary.   

{\bf Numerical modeling}. The $I_c(H)$ dependencies were obtained by solving the sine-Gordon equation:
\begin{equation*}\label{sin}
    \frac{\partial^2 \varphi}{\partial \bar{x}^2} - \frac{\partial^2 \varphi}{\partial \bar{t}^2} - \alpha\frac{\partial \varphi}{\partial \bar{t}} = \sin \varphi.
\end{equation*}
Here $\varphi$ is the Josephson phase difference, $\alpha$ is the quasiparticle damping parameter, the space coordinate $\bar{x} = x/\lambda_J$ is normalized by $\lambda_J$ and time $\bar{t} = \omega_p t$ by the inverse plasma frequency. 

The self-field enters in the boundary condition, 
\begin{equation*}
    \frac{\partial \varphi}{\partial x}\bigg|_{x=0,L_x} = \frac{2\pi}{L_x H_0} \left[H_{e} +  H_{sf}\right], 
\end{equation*}
where $H_{e}$ is the external magnetic field, $H_0$ is the flux quantization field, and $H_{sf}$ is the self-field induced by the asymmetric bias current $I_b(x)$. The self-flux is:
\begin{equation*}
    \Phi_{sf}=L_{sf}I_b \simeq \frac{\lambda_J}{L_x} \frac{\Phi_0}{H_0}H_{sf}.
\end{equation*}
Here, the factor $\lambda_J/L_x$ arises from the exponential decay of the self-flux at a distance $\lambda_J$ from the bias edge. Note that magnetic field required for compensating of self-field is two times smaller, $H_e = -1/2H_{sf}$. This happens because $H_e$ appears in both boundary conditions, $x=0,~L_x$, while $H_{sf}$ only in one (for asymmetric bias).  

Additional details, clarifications and data could be found in the Supplementary information.

\subsection*{Acknowledgements}

\subsection*{Author Contributions}

R.A.H. performed measurements, analyzed data, carried out numerical simulations and participated in writing the manuscript. A.L. participated in transport characterization. T.G. fabricated samples. V.M.K. conceived the project and wrote the manuscript with input from all authors.

\subsection*{Data availability}
Authors confirm that all relevant data are included in the paper and its supplementary information files. Additional data can be provided upon reasonable request.

\subsection*{Competing Interests} The authors declare no competing interests.


\end{document}